\documentstyle[11pt]{article}
\def\be{\begin{equation}}\def\ee{\end{equation}}\def\ni{\noindent}
\def\bb{\begin{thebibliography}{99}\small\addtolength{\itemsep}{-6pt}}
\def\eb{\end{thebibliography}}
\newcommand{\pcf}{{\rm PCF}}\newcommand{\tr}{{\rm tr}\,}
\newcommand{\eqr}[1]{(\ref{#1})}
\def\cN{{\cal N}}\def\e{\hbox{e}}
\def\O{\Omega}\def\lo{\lambda_o}
\def\da{\dagger}\def\a{\alpha}
\begin{document}
\begin{flushright}hep-th/9610147\\
OUTP-96-62P\\October 1996\end{flushright}\vskip2cm
\begin{center}{\LARGE\bf Two-dimensional dynamics of QCD}{\bf 3}\\
\vskip2.5cm {\large\bf B. Rusakov}\\\vskip.5cm{\em Theoretical Physics, 
Oxford University\\1 Keble Road, Oxford, OX1 3NP, UK}\\{\small e-mail: 
rusakov@thphys.ox.ac.uk}\vskip2.5cm {\large\bf Abstract.}\end{center}
Exact loop-variables formulation of pure gauge lattice QCD$_3$ is
derived from the Wilson version of the model. The observation is made
that the resulting model is two-dimensional. This significant feature is
shown to be a unique property of the gauge field. The model is defined on
the infinite genus surface which covers regularly the original
three-dimensional lattice. Similar transformation applied to the principal
chiral field model in two and three dimensions for comparison with QCD. 

\newpage
QCD is believed to be a genuine model of the strong interaction. 
This belief is mainly based on the discovery of asymptotic freedom 
\cite{GW} and on the subsequent success of the perturbative description of
ultra-violet physics. However, the main mystery of the strong 
interactions, its infra-red behavior and, in particular, the confinement
of quarks, remains unexplained because of the lack of a non-perturbative
solution. A number of `mechanisms' of confinement have been suggested.
However, no direct link has been unambiguously established between the
model itself and any particular `mechanism'. Nevertheless, the model is
still analyzable by direct methods. 

The unique, in many respects, non-perturbative formulation of QCD is the
Wilson lattice model \cite{Wils}. The partition function of the pure gauge
model in $D$ dimensions has the form \be Z=\int\prod_ldU_l
\prod_pdU_p\e^{\frac{N}{\lo}\tr(U_p+U_p^\da)}\delta\Big(U_p,\prod_{l
\in p}U_l\Big)\;\;,\label{Z}\ee where $\lo$ is the bare coupling constant, 
$l$ and $p$ denote links and plaquettes of $D$-dimensional lattice, $U_l$
is the unitary matrix ($U(N)$ or $SU(N)$) attached to $l$-th link. In all
the products the link variables are ordered.
 
The gauge-invariant $\delta$-function is\be\delta\Big(U_p,\prod_lU_l\Big)
=\sum_r\chi_r(U_p^\da)\chi_r\Big(\prod_lU_l\Big)\;,\label{d}\ee where $r$ 
is an irreducible representation with character $\chi_r(U)$. 

The only non-zero observables in this model are invariant ordered products
of link variables along {\it closed} loops, such as\be W(C)=
\langle\frac{1}{N}\tr\prod_{l\in C}U_l\rangle\label{wl}\ee ($C$ is the
closed contour). It has been understood that the loop variables are
relevant to solution of the quark confinement problem \cite{Wils,Pol}. In
particular,
as argued in \cite{Wils}, the simplest criterion of confinement is the
area-law behavior of \eqr{wl}: $W(C)\sim\exp(-\sigma A)$, where $A$ is
the area of the `minimal' surface bounded by $C$ and $\sigma$ is the
positive parameter (string tension) to be computed. On a lattice level,
$A$ is the number of plaquettes of a disk bounded by $C$. Since the loop 
variables are only non-zero observables, it is clear that the model
{\it can be} formulated in terms of these variables only. There is a
number of works devoted to the loop dynamics (see review \cite{Mig83} and
references therein). However, this has not led to a realization of
confinement. In contrast, one can argue that if the model is a free theory
in terms of the plaquette variables, then the area law is straightforward.
Such realization requires direct reformulation of the partition function
\eqr{Z} to the loop (plaquette) variables. The latter is the main purpose
of present paper\footnote{See note added to the end of this paper.}.

To rewrite partition function \eqr{Z} in terms of the plaquette variables
only, we have to perform integration over all link variables.

First, we replace \eqr{d} by\be\delta\Big(I,U_p^\da\prod_lU_l\Big)=
\sum_rd_r\chi_r\Big(U_p^\da\prod_lU_l\Big)\;,\label{d1}\ee where 
$d_r=\chi_r(I)$ is the dimension of representation. Such replacement does 
not change \eqr{Z} due to invariance of the action with respect to
similarity transformations $U_p\to\O U_p\O^\da$ and due to
the formula\be\int d\O\chi_r(\O A\O^\da B)=\frac{1}{d_r}\chi_r(A)\chi_r(B)\;
.\label{do}\ee However, there is a considerable difference in applications.
In the case of definition \eqr{d}, the $\O$-integrals are completely
decoupled, while in the case of \eqr{d1}, they play an important role. The
reason we choose the definition \eqr{d1}, though at a first glance it 
seems an unnecessary complication, will soon become clear. Actually, we
will use the freedom to choose the point of insertion of $U_p^\da$ into
the product of $U_l$'s in \eqr{d1}, which means that we mark the point
where the product around the plaquette starts (and ends).
 
Now we change the order of integration in \eqr{Z} and consider the
(topological) integral over link variables \be\int\prod_l^{\cN_1-\cN_0}dU_l
\prod_p^{\cN_2}\delta\Big(U_p,\prod_{l\in p}U_l\Big)\;\;,\label{top}\ee
where $\cN_k$ is the number of $k$-simplexes: $\cN_0$ is the number
of vertices, $\cN_1$ is the number of links etc. The number of independent
integrals is $\cN_1-\cN_0$ due to the gauge invariance. 

In this paper, we only consider the 3d case\footnote{The 2d model is 
exactly solvable \cite{Mig} and its general solution has been found in
\cite{90}. The 4d case will be considered in the next paper \cite{4d}.}.
Due to the Euler theorem for the regular infinite 3d lattice,
\be\cN_0-\cN_1+\cN_2-\cN_3=0\;,\label{e3d}\ee the integral \eqr{top} must 
be proportional to the product of $\cN_3$ $\delta$-functions of $U_p$'s.

\begin{figure}\centering\begin{picture}(200,120)(- 100,- 50)
\thicklines
\put(-80,-19){\line(1,0){100}}\put(-79,-20){\line(0,1){100}}
\put(-80,-21){\line(1,0){100}}\put(-81,-20){\line(0,1){100}}
\put(-80, 80){\line(1,0){100}}\put( 20,-20){\line(0,1){100}}
\put(-20,109){\line(1,0){100}}\put(-20,111){\line(1,0){100}}
\put( 79,10){\line(0,1){100}}\put( 81,10){\line(0,1){100}}
\put( 20,-19){\line(2,1){60}}\put( 20,-21){\line(2,1){60}}
\put(-80,79){\line(2,1){60}}\put(-80,81){\line(2,1){60}}
\put( 20,80){\line(2,1){60}}
\put(-30,-20){\line(-1,1){50}}\put(-55,5){\vector(-1,1){5}}
\put( 30,110){\line(-6,-1){75}}\put(-10,103){\vector(4,1){5}}
\put( 80, 60){\line(-1,-2){33}}\put(65,30){\vector(-1,-2){5}}
\put( 30,110){\line(1,-1){50}}\put(55,85){\vector(1,-1){5}}
\put(-30,-20){\line(6,1){75}}\put(10,-13){\vector(-4,-1){5}}
\put(-80, 30){\line(1,2){33}}\put(-65,60){\vector(1,2){5}}
\put(-20,10){\line(1,0){100}}\put(-20,10){\line(0,1){100}}
\put(-80,-20){\line(2,1){60}}
\put(10,70){\makebox(0,0){A}}\put(-10,20){\makebox(0,0){B}}
\put(20,80){\circle*{5}}\put(-20,10){\circle*{5}}
\end{picture}\caption[x]{\footnotesize Choice of $A$- and $B$-vertices 
and the hexagonal section of a cube.}
\label{fig1}\end{figure}
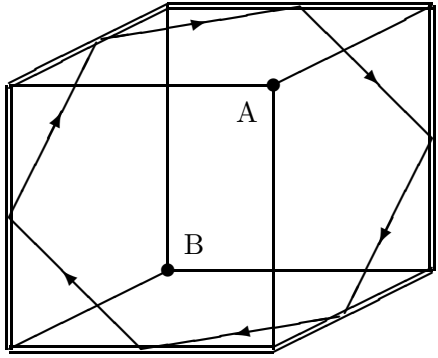

Thus we have to find the set of $\cN_3$ constraints (one for each cube)
self-consistent with the original definition of the model. Let us choose
two vertices, $A$ and $B$, of any cube, that are maximally remote from
each other (Fig.\ref{fig1}), and imply the choice is valid for the
neighboring cubes, i.e. the $A$-vertex is the same for all four cubes it
belongs to, and similarly for $B$. Then, for the rest of the lattice, all 
$A$- and $B$-vertices are uniquely defined. The boundary of any plaquette
passes through either an $A$- or a $B$-vertex and we call the corresponding 
plaquette either a $p_A$- or a $p_B$-plaquette. Now we use our freedom in
placing the $U_p$-matrices into the product around the 
plaquette\footnote{See comments to \eqr{d1}.} so that the product of links
around the $p_A$-plaquettes starts at the vertex $A$, and similarly for
$B$. Then, the invariant ordered product of three $U_{p_A}$'s of each cube
is equal to one of $U_{p_B}$'s of the same cube, i.e., 
$$\chi_r\Big(\prod_{p_A}U_{p_A}\Big)=\chi_r\Big(\prod_{p_B}U_{p_B}\Big)
,$$ and both are equal to the invariant ordered product of $U_l$'s, where
$l$'s form the closed loop separating $A$ and $B$ on the cube surface
(double line in Fig.\ref{fig1}). Then, the sought-for constraint is
expressed by the formula:
\be\delta\Big(\prod_{p_A}U_{p_A},\prod_{p_B}U_{p_B}\Big)=\sum_r
\chi_r\Big(\prod_{p_A}U_{p_A}^\da\Big)\chi_r\Big(\prod_{p_B}U_{p_B}\Big)
\;,\label{dp}\ee where the $p_A$- and $p_B$-plaquettes belong to the same 
cube. It is clear that all the constraints are independent. Thus, the
integral \eqr{top} is equal (up to a possible constant factor) to the
product of $\delta$-functions \eqr{dp} taken over all cubes $c$. This has
to be substituted into \eqr{Z}:
 
\be Z=\int\prod_pdU_p\e^{\frac{N}{\lo}\tr(U_p+U_p^\da)}\prod_c
\delta\Big(\prod_{p_A\in c}U_{p_A},\prod_{p_B\in c}U_{p_B}\Big)
\;.\label{Z1}\ee

Thus, we formulated the original model \eqr{Z} in terms of the plaquette
variables only. 

We notice now that this formulation has an interesting feature. Namely, 
\eqr{Z1} is the partition function of a certain {\bf two-dimensional}
model. To see this, let us consider the section of a cube by
hexagon, as shown at Fig.\ref{fig1}. Using the one-to-one correspondence
between the plaquettes of original lattice and the links of the hexagon,
we replace the $U_p$-variables in \eqr{Z1} by (new) $U_l$'s where $l$ is
the corresponding link of the hexagon. 

\def\le{\left(\begin{array}{ll}&\\&\end{array}\right.}
\def\ri{\left.\begin{array}{ll}&\\&\end{array}\right)}
\def\rob{\begin{picture}(15,5)(-15,-5)
\put(14.5,-1.7){\makebox(0,0){.}}\put(-14.5,-1.7){\makebox(0,0){.}}
\put(14,-2.1){\makebox(0,0){.}}\put(-14,-2.1){\makebox(0,0){.}}
\put(13.5,-2.4){\makebox(0,0){.}}\put(-13.5,-2.4){\makebox(0,0){.}}
\put(13,-2.6){\makebox(0,0){.}}\put(-13,-2.6){\makebox(0,0){.}}
\put(12,-2.85){\makebox(0,0){.}}\put(-12,-2.85){\makebox(0,0){.}}
\put(11,-3.05){\makebox(0,0){.}}\put(-11,-3.05){\makebox(0,0){.}}
\put(10,-3.2){\makebox(0,0){.}}\put(-10,-3.2){\makebox(0,0){.}}
\put(9,-3.33){\makebox(0,0){.}}\put(-9,-3.33){\makebox(0,0){.}}
\put(8,-3.45){\makebox(0,0){.}}\put(-8,-3.45){\makebox(0,0){.}}
\put(7,-3.55){\makebox(0,0){.}}\put(-7,-3.55){\makebox(0,0){.}}
\put(6,-3.63){\makebox(0,0){.}}\put(-6,-3.63){\makebox(0,0){.}}
\put(5,-3.70){\makebox(0,0){.}}\put(-5,-3.70){\makebox(0,0){.}}
\put(4,-3.76){\makebox(0,0){.}}\put(-4,-3.76){\makebox(0,0){.}}
\put(3,-3.81){\makebox(0,0){.}}\put(-3,-3.81){\makebox(0,0){.}}
\put(2,-3.85){\makebox(0,0){.}}\put(-2,-3.85){\makebox(0,0){.}}
\put(1,-3.88){\makebox(0,0){.}}\put(-1,-3.88){\makebox(0,0){.}}
\put(0.5,-3.9){\makebox(0,0){.}}\put(-0.5,-3.9){\makebox(0,0){.}}
\put(0.2,-3.91){\makebox(0,0){.}}\put(-0.2,-3.91){\makebox(0,0){.}}
\end{picture}}

\def\rot{\begin{picture}(15,5)(-15,-5)
\put(14.5,1.7){\makebox(0,0){.}}\put(-14.5,1.7){\makebox(0,0){.}}
\put(14,2.1){\makebox(0,0){.}}\put(-14,2.1){\makebox(0,0){.}}
\put(13.5,2.4){\makebox(0,0){.}}\put(-13.5,2.4){\makebox(0,0){.}}
\put(13,2.6){\makebox(0,0){.}}\put(-13,2.6){\makebox(0,0){.}}
\put(12,2.85){\makebox(0,0){.}}\put(-12,2.85){\makebox(0,0){.}}
\put(11,3.05){\makebox(0,0){.}}\put(-11,3.05){\makebox(0,0){.}}
\put(10,3.2){\makebox(0,0){.}}\put(-10,3.2){\makebox(0,0){.}}
\put(9,3.33){\makebox(0,0){.}}\put(-9,3.33){\makebox(0,0){.}}
\put(8,3.45){\makebox(0,0){.}}\put(-8,3.45){\makebox(0,0){.}}
\put(7,3.55){\makebox(0,0){.}}\put(-7,3.55){\makebox(0,0){.}}
\put(6,3.63){\makebox(0,0){.}}\put(-6,3.63){\makebox(0,0){.}}
\put(5,3.70){\makebox(0,0){.}}\put(-5,3.70){\makebox(0,0){.}}
\put(4,3.76){\makebox(0,0){.}}\put(-4,3.76){\makebox(0,0){.}}
\put(3,3.81){\makebox(0,0){.}}\put(-3,3.81){\makebox(0,0){.}}
\put(2,3.85){\makebox(0,0){.}}\put(-2,3.85){\makebox(0,0){.}}
\put(1,3.88){\makebox(0,0){.}}\put(-1,3.88){\makebox(0,0){.}}
\put(0.5,3.9){\makebox(0,0){.}}\put(-0.5,3.9){\makebox(0,0){.}}
\put(0.2,3.91){\makebox(0,0){.}}\put(-0.2,3.91){\makebox(0,0){.}}
\end{picture}}

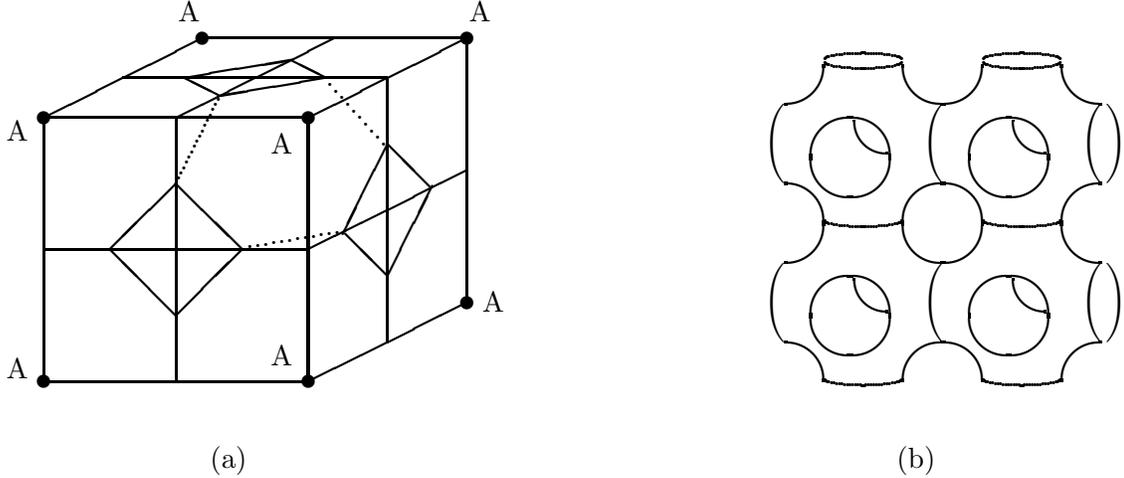
\begin{figure}
\centering\begin{picture}(200,170)(-100,-50)\thicklines
\put(-200,-20){\line(1,0){100}}\put(-200,-20){\line(0,1){100}}
\put(-200, 80){\line(1,0){100}}\put(-100,-20){\line(0,1){100}}
\put(-150,-20){\line(0,1){100}}\put(-200, 30){\line(1,0){100}}
\put(-140,110){\line(1,0){100}}\put(-40,10){\line(0,1){100}}
\put(-170,95){\line(1,0){100}}\put(-70,-5){\line(0,1){100}}
\put(-100,-20){\line(2,1){60}}\put(-200,80){\line(2,1){60}}
\put(-100,80){\line(2,1){60}}\put(-150,80){\line(2,1){60}}
\put(-100,30){\line(2,1){60}}
\put(-150,5){\line(1,1){25}}\put(-150,5){\line(-1,1){25}}
\put(-150,55){\line(1,-1){25}}\put(-150,55){\line(-1,-1){25}}
\multiput(-150,55)(1,2){17}{\makebox(0,0){.}}
\put(-70,20){\line(-1,1){16}}\put(-70,20){\line(1,2){17}}
\put(-70,70){\line(1,-1){16}}\put(-70,70){\line(-1,-2){17}}
\multiput(-86,37)(-6,-1){7}{\makebox(0,0){.}}
\multiput(-89,{36.5})(-6,-1){6}{\makebox(0,0){.}}
\put(-147,95){\line(2,-1){13}}\put(-147,95){\line(6,1){41}}
\put(-93,95){\line(-2,1){13}}\put(-93,95){\line(-6,-1){41}}
\multiput(-94,95)(9,-10){3}{\makebox(0,0){.}}
\multiput(-91.75,92.5)(9,-10){3}{\makebox(0,0){.}}
\multiput(-89.5,90)(9,-10){3}{\makebox(0,0){.}}
\multiput(-87.25,87.5)(9,-10){2}{\makebox(0,0){.}}
\put(-210,-15){\makebox(0,0){A}}\put(-200,-20){\circle*{5}}
\put(-110,-10){\makebox(0,0){A}}\put(-100,-20){\circle*{5}}
\put(-210,75){\makebox(0,0){A}}\put(-200,80){\circle*{5}}
\put(-110,70){\makebox(0,0){A}}\put(-100,80){\circle*{5}}
\put(-145,120){\makebox(0,0){A}}\put(-140,110){\circle*{5}}
\put(-35,120){\makebox(0,0){A}}\put(-40,110){\circle*{5}}
\put(-30,10){\makebox(0,0){A}}\put(-40,10){\circle*{5}}
\put(80,-20){\oval(30,30)[tr]}\put(140,-20){\oval(30,30)[t]}
\put(200,-20){\oval(30,30)[tl]}\put(80,40){\oval(30,30)[r]}
\put(140,40){\oval(30,30)}\put(200,40){\oval(30,30)[l]}
\put(80,100){\oval(30,30)[br]}\put(140,100){\oval(30,30)[b]}
\put(200,100){\oval(30,30)[bl]}

\put(105,65){\oval(30,30)}\put(119,79){\oval(25,25)[bl]}
\put(105,5){\oval(30,30)}\put(119,19){\oval(25,25)[bl]}
\put(165,65){\oval(30,30)}\put(179,79){\oval(25,25)[bl]}
\put(165,5){\oval(30,30)}\put(179,19){\oval(25,25)[bl]}
\put(102.5,100){\makebox(0,0){\rob}}\put(102.5,98){\makebox(0,0){\rot}}
\put(162.5,100){\makebox(0,0){\rob}}\put(162.5,98){\makebox(0,0){\rot}}
\put(102.5,40){\makebox(0,0){\rob}}\put(162.5,40){\makebox(0,0){\rob}}
\put(102.5,-20){\makebox(0,0){\rob}}\put(162.5,-20){\makebox(0,0){\rob}}
\put(207,70){\makebox(0,0){$\le$}}\put(195,70){\makebox(0,0){$\ri$}}
\put(207,10){\makebox(0,0){$\le$}}\put(195,10){\makebox(0,0){$\ri$}}
\put(87,70){\makebox(0,0){$\le$}}\put(87,10){\makebox(0,0){$\le$}}
\put(147,70){\makebox(0,0){$\le$}}\put(147,10){\makebox(0,0){$\le$}}
\put(-130,-50){\makebox(0,0){(a)}}\put( 130,-50){\makebox(0,0){(b)}}
\end{picture}\caption[x]{\footnotesize (a) The 8-cube's fragment of the 
lattice. $A$-vertices are the same as in Fig.1. Hexagons (sections of
cubes) form the 2d lattice (not all the lines are shown in order to not
disturb the view). (b) The fragment of the resulting (smoothed) surface
(corresponds to 32 cubes of original lattice).}\label{fig2}\end{figure}

The hexagons form the 2d lattice (Fig.\ref{fig2}). The partition function 
\eqr{Z1}, in terms of this lattice, takes the form\be Z=\int\prod_l
dU_l\e^{\frac{N}{\lo}
\tr(U_l+U_l^\da)}\prod_h\delta\Big(\prod_{l_1\in h}U_{l_1},\prod_{l_2\in h}
U_{l_2}\Big)\;,\label{Z2}\ee where $l$ and $h$ denote links and hexagons of 
the 2d lattice. The order of links in any hexagon (argument of
$\delta$-function) is $l_1l_2l_1l_2l_1l_2$ (at Fig.2(a), the $l_2$-type 
links correspond to the dotted lines). The fragment of the resulting 2d
lattice is shown at Fig.2(b). The whole lattice is easy to imagine -- it 
is obtained from the regular 3d lattice (the lattice spacing is doubled
with respect to the original 3d lattice) by replacing links by the tubes
(handles) and vertices by the smooth connections of tubes. The surface 
separates all $A$-vertices  from all $B$-vertices. The two sets, being
interior and exterior to the surface, are identical to each other. The
genus $g$ of the surface is proportional to $\cN_3$, namely,
$g=\frac{3}{8}\cN_3$. 

Thus, we started with the three-dimensional field and discovered that,
actually, interaction takes place on a certain surface. In our view, the
picture becomes more clear in terms of the $r$ field (where $r$ is the
$N$-component highest weight vector of representation; see definition
\eqr{d}). Each particular sum over $r$ is attached to the particular
hexagon and we can consider \eqr{Z2} as the partition function of 
statistical model of the $N$-component 2d field $r$ (\eqr{Z2} is the sum
over all $r$-configurations). 

It is clear from the above that the reduction described is a unique
property of the gauge field and takes place due to the gauge invariance of
the original model \eqr{Z}.  However, it is still interesting to see
what happens in other models if similar change of variables is performed. 

We consider the principal chiral field (PCF) model. Its partition function
has the form\be Z_{\pcf}=\int\prod_idU_i\prod_ldU_l
\e^{\frac{N}{\lo}\tr(U_l+U_l^\da)}\delta(U_{l=\{ij\}},U_iU_j^\da)\;\;,
\label{pcf}\ee where $i,j$ are the vertex labels, $l$ denote links and
$l=\{ij\}$ denotes the link between two neighboring vertices $i$ and $j$.

Now our aim is to formulate this model in terms of link variables only,
i.e., to perform integration over vertices. Due to the Euler theorem,
$\cN_0-\cN_1+\cN_2=0$, the topological integral over vertices produces
$\cN_2$ $\delta$-functions (or $\cN_2-\cN_3+...$ $\delta$-functions, 
if dimension is higher than two). The corresponding condition for the
plaquette is, obviously,\be\delta\Big(I,\prod_{l\in p}U_l\Big)\ee (this
condition is valid for any loop, but we still keep the notation $p$).
Thus,\be Z_{\pcf}=\int\prod_ldU_l\e^{\frac{N}{\lo}\tr(U_l+U_l^\da)}\prod_p
\delta\Big(I,\prod_{l\in p}U_l\Big)\;.\label{zpcf}\ee Here, $p$ counts all
independent loops, where `independence' of the loop means an independence
of the corresponding constraint. In particular, in the 2d PCF all the
loops are independent (unless this 2d is the compact surface, in which
case only one loop has to be excluded). 

Let us consider PCF on the surface shown at Fig.2, in order to
compare with QCD$_3$. There are conditions $\delta\Big(I,\prod_lU_l\Big)$
at the non-contractible loops (the non-trivial cycles). Field $r$
corresponding to such a loop is attached to the section of the tube, i.e.,
there is a three-dimensional interaction inside the handles, in sharp
contrast to QCD.

If in \eqr{zpcf} we replace the $\delta$-function by the heat kernel
\be f(\a,U)=\sum_rd_r\e^{-\frac{\a}{N}C_2(r)}\chi_r(U)\label{hk}\ee
($C_2(r)$ is the second Casimir eigenvalue) and introduce a different
couplings, $\a_{hex}$ for the hexagons, and $\a_{tub}$ for the sections of
the tubes, then $Z_\pcf$ is restored as the $\a_{hex},\a_{tub}\to 0$ 
limit of the result, while QCD$_3$ partition function \eqr{Z2} is
the $\a_{hex}\to 0$, $\a_{tub}\to\infty$ limit.

There are several comments in order. The first one is the technical
comment on use of \eqr{d1} and \eqr{do} instead of \eqr{d}. It has been
understood already in QCD$_2$ \cite{90} that the $\O$-degrees of freedom,
though they are hidden in the  model, play an important role. In QCD$_2$, 
they act by means of the formula \eqr{do} which pulls the powers of $d_r$
to denominator and, thus, the $1/N$-expansion of the QCD$_2$ partition
function \cite{90} can be done \cite{GT}. Here, in three dimensions, the
use of the $\O$-degrees of freedom allowed us to decrease the effective
dimension of the system. For reader familiar with the hermitian matrix
models we remind that the similar situation takes place there. In all
these cases, the use of the $\O$-variables has led to the kind of a
`stringy' description. It is worth to remind that it is these variables,
which are precisely absent in any abelian model. 

The important question is what happens with this picture in continuum
limit. The loop variables are only non-zero observables in continuum
theory as well. Therefore, the principal possibility to formulate the
continuum model completely in terms of the loop variables still exists
\cite{Pol}. From the present consideration we see that the change of
variables from links to plaquettes can be done at any lattice scale.
Suppose we approach the continuum limit by refinement of the 3d lattice so
that the model is restored at each finer level. This should be achieved by 
the appropriate tuning of the coupling constant. Then at each level we can
restore the described 2d picture with the coupling constant corresponding
to the current level of refinement. However, the 2d lattice undergoes the
infinite change of genus under refinement of the original 3d lattice. 
The quantity which survives the refinement is the ratio $\cN_3/g=8/3$,
where $\cN_3$ is the volume of the original lattice. 

\vskip.3cm
\ni{\bf Note added:} After submission of this work, I was informed
of the related, pioneering work \cite{HB} on field strength and plaquette
variable formulation of gauge theory. The relation between this early work
and present paper is currently under investigation.

\bb
\bibitem{GW} D.Gross and F.Wilczek, Phys. Rev. {\bf D8} (1973) 3633 \&
                                    Phys. Rev. {\bf D9} (1974) 980. 
\bibitem{Wils} K.Wilson, Phys. Rev. {\bf D10} (1974) 2445.
\bibitem{Pol} A.Polyakov, {\em Gauge Fields and Strings}, Harwood, 1987.
\bibitem{Mig83} A.Migdal, Phys. Rep. {\bf 102} (1983) 199.
\bibitem{Mig} A.Migdal, ZhETF {\bf 69} (1975) 810.
\bibitem{90} B.Rusakov, Mod. Phys. Lett. {\bf A5} (1990) 693.
\bibitem{4d} B.Rusakov, {\em in preparation}.
\bibitem{GT} D.Gross, Nucl. Phys. {\bf B400} (1993) 161;
             D.Gross and W.Taylor, Nucl. Phys. {\bf B400} (1993) 181 \&
                                   Nucl. Phys. {\bf B403} (1993) 395.
\bibitem{HB} M.B.Halpern, Phys. Rev. {\bf D19} (1979) 517; G.G.Batrouni,
Nucl. Phys. {\bf B208} (1982) 12; 467; G.G.Batrouni and M.B.Halpern,
Phys. Rev. {\bf D30} (1984) 1775; 1782.
\eb\end{document}